\documentclass[aps,prb,twocolumn,showpacs,preprintnumbers,amsmath,amssymb]{revtex4}
\usepackage{graphicx}
\usepackage{bm}
\usepackage{epsfig}

\begin{document}

\newcommand{\bec}{\begin{center}}
\newcommand{\ec}{\end{center}}
\newcommand{\be}{\begin{equation}}
\newcommand{\ee}{\end{equation}}
\newcommand{\beqn}{\begin{eqnarray}}
\newcommand{\eeqn}{\end{eqnarray}}
\newcommand{\bet}{\begin{table}}
\newcommand{\ent}{\end{table}}
\newcommand{\bib}{\bibitem}

\title{
Self-organized transient facilitated atomic transport in Pt/Al(111)
}

\author{P. S\"ule} 
  \address{Research Institute for Technical Physics and Material Science,\\
Konkoly Thege u. 29-33, Budapest, Hungary,sule@mfa.kfki.hu,www.mfa.kfki.hu/$\sim$sule,\\
}

\date{\today}


\begin{abstract}
During the course of atomic transport in a host material, impurity atoms need to surmount 
an energy barrier driven by thermodynamic bias or at ultra-low temperatures
by quantum tunneling.
In the present article we demonstrate using atomistic simulations that at ultra-low temperature
transient inter-layer atomic transport is also possible without tunneling
when the Pt/Al(111) impurity/host system self-organizes itself spontaneously into
an intermixed configuration.
No such extremely fast athermal concerted process has been reported before at ultra low temperatures.
The outlined novel transient atomic exchange mechanism could be of general validity.
We find that 
the source of ultra-low temperature heavy particle barrier crossing is intrinsic and no external
bias is necessary for atomic intermixing and surface alloying in Pt/Al although
the dynamic barrier height is few eV.
The mechanism is driven by the local thermalization of the Al(111) surface
in a self-organized manner arranged spontaneously by the system
without any external stimulus.
The core of the short lived thermalized region reaches the local temperature of $\sim 1000$ K 
(including few tens of Al atoms)
while the average temperature
of the simulation cell is $\sim 3$ K.
The transient facilitated intermixing process also takes place with repulsive
impurity-host interaction potential leading to negative atomic mobility
hence the atomic injection is largely independent of the
strength of the impurity-surface interaction.
We predict that similar exotic behaviour is possible in other materials as well.


\end{abstract}
\maketitle

\section{Introduction}

  The spontaneous formation of self-organized nanoscale structures
has attracted much attention in recent years due to its potential application
in the fabrication of nanodevices \cite{Schukin,He}.
 The understanding of atomistic processes which lead to the formation of
nanostructures has been one of the main focuses of research in materials
science \cite{Schukin,Teichert}.

 The self-assembly induced atomic movements towards an ordered structure can be understood as thermally activated processes 
driven by the thermodynamic bias of the system \cite{Schukin,Philibert}.
Also, concerted atomic transport processes during self-organization
such as adatom nucleation via detechment and attachment processes at step edges and thin film growth and processing, however, often lead to abrupt surface alloying and
intermixing \cite{Michely,Brune,Tersoff,Buchanan}.
These processes proceed via 
atomic site exchanges within the topmost atomic layer \cite{siteexchange,Bulou,Ehrlich,Ferrando}.

 Ultrafast diffusional dynamics can be studied by classical molecular dynamics (MD) simulations at the atomistic level \cite{Bulou,Ferrando,Allen,Levy,Luedtke,Nordlund_Nature}.
Recently it has been shown by MD studies in accordance with experimental results that
 under externally forced conditions,
 transient enhanced intermixing of heavier impurities could occur in bulk materials \cite{Sule_JAP07,Sule_PRB05}.
In the absence of considerable external load of perturbation, such as 
during atomic deposition ultrafast intermixing and surface alloying can also be induced \cite{Tersoff,Buchanan,Fichthorn,Haftel1,Sprague}.
Moreover, the most recently it has also been found that the bulk mobility of atomic metallic clusters
could also be enhanced leading to ballistic burrowing in Al and in Ti \cite{Sule_cluster}.

 These are interesting results beacuse it is widely accepted that enhanced diffusion
occurs the mostly on solid surfaces e.g.
 when the barrier of atomic transport $\Delta E \le k_B T$, where $k_B $ and $T$ are the Boltzmann constant and the temperature, respectively, superdiffusion occurs, that is the nearly dissipationless atomic transport with transient atomic jumps 
(random walk, Levy flight) \cite{Michely,Brune,Levy,Luedtke,Sule_SUCI}.
In the topmost layer fast atomic exchange processes with long jumps have also been
reported which thought to be driven, however, by thermodynamic forces \cite{Bulou,Ehrlich,Ferrando}.
In the bulk, non-Arrhenius (athermal) atomic (not necessarily transient) transport has been studied mostly under
nonequilibrium conditions, such as during ion-implantation \cite{Nordlund_Nature},
in driven-alloys or mechanical alloying \cite{Martin,Lund},
by mechanical force biased chemical reactions
\cite{mechanic} or
using shock-induced alloying \cite{shock}. 

 In the bulk
 athermal rates can only be accomplished
via under barrier atomic jumps called quantum tunneling \cite{Philibert,Kramer}. 
This can be done mostly for light particles at ultra-low temperatures.
 Quantum diffusion of H has been studied in detail on solid surfaces \cite{Honsurf}.
In the bulk,
only few light elements show ultrafast interstitial diffusion 
\cite{Philibert,Abrasonis,Marx}.
However, the quantum tunneling diffusion of heavy adatoms on various substrate surfaces have also been observed recently
\cite{Bulou2,Chan}.
  Quantum diffusion (QD) have not been observed yet for heavy elements in the bulk 
although
the de Broglie wavelength could be in the range of tunneling distance which allows
QD on the surface \cite{Bulou2}.
The most recently reactive diffusion dynamics has been interpreted
as a superdiffusive process during the front propagation of interfaces \cite{Brockmann}
which could be the first (though theoretical) finding that transient diffusive
atomic transport takes place in the bulk.

 We would like to present classical MD results which suggest that transient rates could also be
occurred without tunneling at ultra-low temperature via a peculiar mechanism during surface alloying.
The employed semiempirical approach is validated by {\em ab initio}
density functional calculations.
The explored new transient atomic exchange process is driven in a self-organized manner:
the impurity/host system spontaneously reorganizes itself in such a way that
abrupt surface alloying takes place at an ultra-low temperature.
We also find that impurity induced local thermalization occurs at $\sim 0$ K external temperature 
although no forced condition has been applied.
Such a mechanism has not been observed yet although
is likely to be of general validity.
The local thermalization of the
substrate facilitates atomic injection while the overall temperature of the
substrate remains very low (few K).
A surprising consequence is that repulsive intermixing (and negative transient atomic mobility)  could also occur theroretically,
the process is not sensitive to the strength of the Al-Pt interaction.

\section{The simulation approach}

 Classical tight-binding molecular dynamics simulations \cite{CR} were used to simulate 
soft landing and vapor deposition of Pt atoms on Al(111) substrate at $\sim 0$ K
using the PARCAS code \cite{Nordlund_ref} which has been used
for the study of various atomic transport phenomena
in the last few years \cite{Sule_PRB05,Nordlund_ref}.
We also employ first principles calculations to validate our heteronuclear potential
(details will be given later on).

 Although we carry out simulations at $\sim 0$ K, we find a substantial local heating
up in a local surface region of Al, hence the correct dissipation of the
emerged heat should be handled via using temperature control.
A variable timestep
and the Berendsen temperature control is used at the cell border 
\cite{Allen,Frenkel,Berendsen}.
The simulation uses the Gear's predictor-corrector algorithm to calculate
atomic trajectories \cite{Allen}.
The maximum time step of $0.05$ fs is used during the operation of the multiple time st
ep algorithm
\cite{Frenkel}.
The system couples to a heat bath via the damping constant
to maintain constant temperature conditions and the thermal equilibrium of the entire
system \cite{Berendsen}.
The time constant for temperature control is chosen to be $\tau=70$ fs, where $\tau$ is
 a characteristic relaxation time to be adjusted \cite{Frenkel,Berendsen}.
The Berendsen temperature control has successfully been used                           for nonequilibrium systems, such as occur during ion-bombardment of various            materials \cite{Nordlund_Nature,Sule_JAP07,Sule_PRB05,Sule_SUCI,Nordlund_ref,PARCAS}.
Further details are given in ref. \cite{Nordlund_ref,PARCAS} and details specific to the current system in recent
communications \cite{Sule_JAP07,Sule_PRB05,Sule_NIMB04}.

  For simulating deposition it is appropriate to use temperature control
at the cell borders.
This is because it is physically correct that potential energy becomes
kinetic energy on impact, i.e. heats the lattice. This heating should
be allowed to dissipate naturally, which means temperature control should not
be used at the impact point.
Periodic boundary conditions are imposed laterarily.
The observed anomalous transport processes are also observed without
periodic boundary conditions and Berendsen temperature control.
Further details are given in \cite{PARCAS} and details specific to the current system in recent
communications \cite{Sule_PRB05,Sule_NIMB04}.

  The top of the simulation cell is left free (the free surface) for
the deposition of Pt atoms.
The bottom layers
are held fixed in order to avoid the rotation of the cell.
Since the z direction is open, rotation could start around the z axis.
The bottom layer fixation is also required to prevent
the translation of the cell.

  The size of the simulation cell is $80 \times 80 \times 42$ $\hbox{\AA}^3$ including
 $16128$ atoms (with a fcc lattice).
The simulation uses the Gear's predictor-corrector algorithm to calculate
atomic trajectories \cite{Frenkel}.
The maximum time step of $0.05$ fs is used during the operation of the multiple time step algorythm
\cite{Frenkel}.
$15$ active MLs are supported on $3$ fixed bottom monolayers (MLs).
We find no dependence of the anomalous atomic transport properties of the deposited atoms on the
finite size of the simulation cell.
Finite size effects do not play a role in the appearance of the anomalous transport
of Pt in Al (the variation of the cell size does not influence the intermixing process
down to cell sizes including few hundreds of atoms).
Deposited atoms were initialized normal to the (111) surface with randomly
selected lateral positions $4-5$ $\hbox{\AA}$ above the surface with nearly
zero velocity.
 The initial kinetic energy of the deposited particles 
in the case of ultrasoft landing is nearly zero eV.
In order to make a statistics of impact events we generated $100$ events with randomly
varied impact positions.
 The conservation of the total energy is maintained during the simulations.

\subsection{The interaction potential}

 We use the many-body 
tight-binding second-moment approximation (TB-SMA) interaction potential to describe interatomic interactions \cite{CR}.
Using the Cleri-Rosato (CR) parameterization of the TB-SMA potential we consider the interaction between two atoms and the interaction
with their local environment.
\begin{figure}[hbtp]
\begin{center}
\includegraphics*[height=5cm,width=6cm,angle=0.]{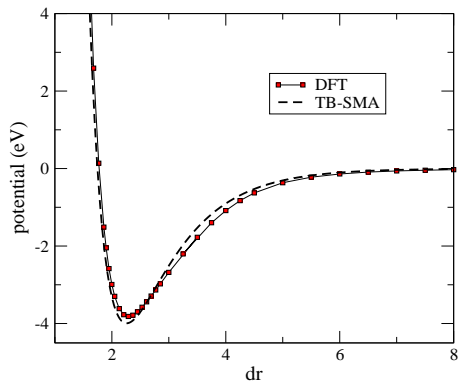}
\caption[]{
The crosspotential energy (eV) for the Al-Pt dimer as a function of
the interatomic distance ($\hbox{\AA}$) obtained by
the {\em ab initio} PBE/DFT method. For comparison the interpolated semiempirical potential
(TB-SMA) is also shown calculated for the Al-Pt dimer.
}
\label{potential}
\end{center}
\end{figure}

 The TB-SMA potential is formally analogous to the embedded atomic method (EAM, \cite{EAM}) 
formalism, e.g.
the potential energy of an atom is given as a sum of repulsive pair potentials for the
neighboring atoms (usually for the first or second neighbors and a cutoff is imposed out of
this
region) and an embedding energy that is a function of the local electron density
given as follows \cite{EAM},
\be
E_{tot}= \frac{1}{2}\sum_{ij} V(r_{ij}) +\sum_i F[\rho_{i}],
\label{eq1}
\ee
where $r_{ij}$ is the distance between atoms $i$ and $j$.
There are many functional forms are available for the density $\rho_{i}$ and for the 
embedding
function $F[\rho_{i}]$ \cite{EAM}.
In the code PARCAS \cite{Nordlund_ref} the forces have been calculated
using a built-in functional derivative of Eq. (1).
We utilize EAM functional forms in the code for $F[\rho_{i}]$ and for the density $\rho$
similar to that given in refs. \cite{EAM,Haftel}.
The EAM routine in the code employs a cubic spline interpolation for the evaluation
of the EAM potentials and their derivatives (forces) starting from various kind of input 
potentials
given in discrete points as a function of $r_{ij}$ (the number of points per functions
is $5000$ in this study).

 Within the TB-SMA we use no explicit dependence on $\rho_{i}$. 
The attractive part of the potential reads,
\be
 F^i(r_{ij})=-\biggm[ \sum_{j, r_{ij} < r_c} \xi^2 exp \biggm[-2q \biggm(\frac{r_{ij}}{r_0}
-1 \biggm) \biggm] \biggm]^{1/2},
\ee
where $r_c$ is the cutoff radius of the interaction and $r_0$ is the first neighbor distance (atomic size parameter).

The repulsive term is a Born-Mayer type phenomenological core-repulsion term:
\be
V^i(r_{ij})=A \sum_{j, r_{ij}<r_c} exp \biggm[-p \biggm(\frac{r_{ij}}{r_0}-1 \biggm) \biggm].
\ee
The parameters ($\xi, q, A, p, r_0$) are fitted to experimental values of the cohesive energy,
the lattice parameter, the bulk modulus and the elastic constants $c_{11}$, $c_{12}$ and $c_{44}$ \cite{CR}
and which are given in Table 1.
The summation over $j$ is extended up to fifth neighbors for fcc structures \cite{CR}.
The cutoff radius $r_c$ is taken as the third neighbor distance for all the interactions.
We tested the Al-Al and the Al-Pt potential at cutoff radius with larger neighbor distances and
found no considerable change in the results.
This type of a potential gives a very good description of lattice vacancies, including migration
properties and a reasonable description of solid surfaces and melting \cite{CR}.
The CR potential correctly provides the adatom binding and dimerization energies
\cite{Sule_SUCI}.

  Recently it has also been shown, that the CR potential remarkably well describes
diffusion in liquid Al \cite{Li2,Alemany} and energetic deposition of
Al clusters on Al \cite{Kang}.
For the Al-Pt crosspotential of substrate atoms and Pt we employ an interpolation scheme which has widely been used in the literature \cite{Bulou,Ferrando,Sule_PRB05,Sule_NIMB04,Meyerheim,Sule_SUCI}.
The Al-Pt potential provides a reasonable melting point 
and heat of alloying for the AlPt alloy \cite{Sule_SUCI}.

\begin{table}
\caption
{
The Cleri-Rosato parameters \cite{CR} used in the tight binding potential (TB-SMA) give
n in Eqs
. (1)-(2) \cite{CR}
The parameters of the crosspotential have been obtained
as follows using an interpolation scheme \cite{AB}:
For the preexponentials $\xi$ and $A$ we used the
harmonic mean $A_{AlPt}=(A_{Al} \times A_{Pt})^{1/2}$
($\xi$ has been fitted to the heat of mixing of the AlPt alloy phase,
see details in ref. \cite{Sule_NIMB04}),
for $q$ and $p$ we use the geometrical averages:
$q_{AlPt}=(q_{Al} + q_{Pt})/2$. The first neighbor distance of the
Al-Pt potential is given also as a geometrical mean of $r_0=(r_0^{Pt}+r_0^{Al})/2$.
}
\begin{tabular}{cccccc}
& $\xi$ & q & A & p & $r_0$  \\
\hline \hline
 Al  & 1.316  & 4.516  & 0.122 & 8.612 & 2.87 \\
 Pt  & 2.695 & 4.004  & 0.298 & 10.612 & 2.78   \\
 Al-Pt & 2.7 & 3.258 & 0.191  & 9.612 & 2.83   \\
\hline \hline
\end{tabular}
\end{table}

 In order to check the accuracy of the employed interpolated crosspotential,
 the crosspotential energy has also been calculated for the Al-Pt dimer
  using {\em ab initio} local spin density functional calculations \cite{G03} together 
with a quadratic convergence self-consistent field method.
The G03 code is well suited for molecular calculations, hence
it can be used for checking pair-potentials.
The Kohn-Sham equations (based on density functional theory, DFT) \cite{KS} are solved in an atom centered Gaussian basis set and the core electrons
are described by effective core potentials
(using the LANL2DZ basis set) \cite{basis}
and
we used the Perwed-Burke-Ernzerhof (PBE) gradient corrected exchange-correlation potential \cite{PBE}.
First principles calculations based on
density functional theory (DFT) have been applied in various fields
in the last few years \cite{DFT_Sule}.

The obtained profile is plotted in Fig. \ref{potential}
together with our interpolated semiempirical many-body TB-SMA potential for the Al-Pt dimer.
We find that our interpolated TB-SMA potential when calculated for the Al-Pt dimer
matches reasonably well the ab initio one hence we are convinced
that the TB-SMA model accurately describes the heteronuclear
interaction in the Al-Pt dimer.
We assume that this dimer potential is transferable for
those cases when the Pt atom is embedded in Al.
This can be done because, as we outlined above, the interpolated
Al-Pt potential properly reproduces the available experimental results
for the Al-Pt alloy.
\begin{figure}[hbtp]
\begin{center}
\includegraphics*[height=3.2cm,width=4.2cm,angle=0.]{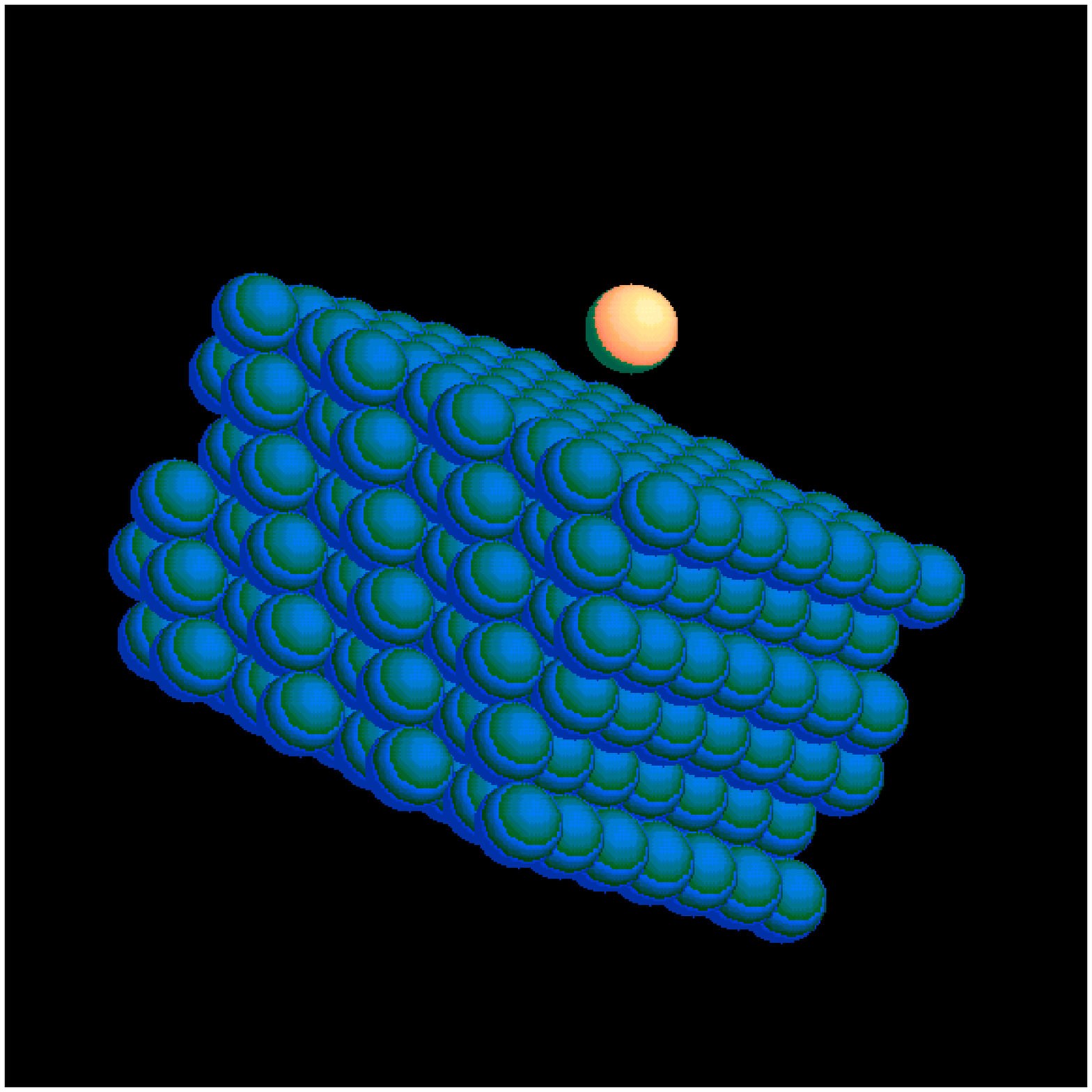}
\includegraphics*[height=3.2cm,width=4.2cm,angle=0.]{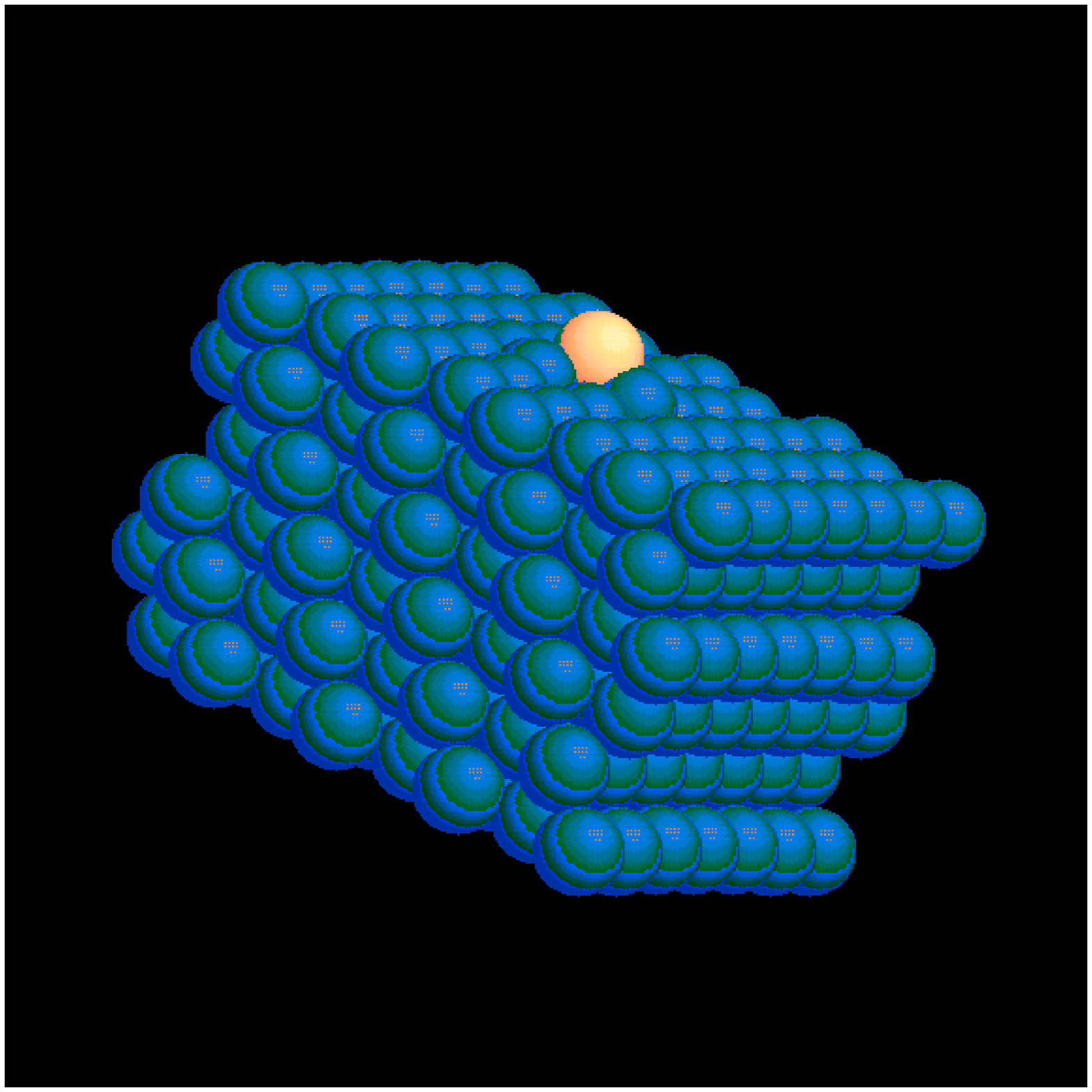}
\includegraphics*[height=3.2cm,width=4.2cm,angle=0.]{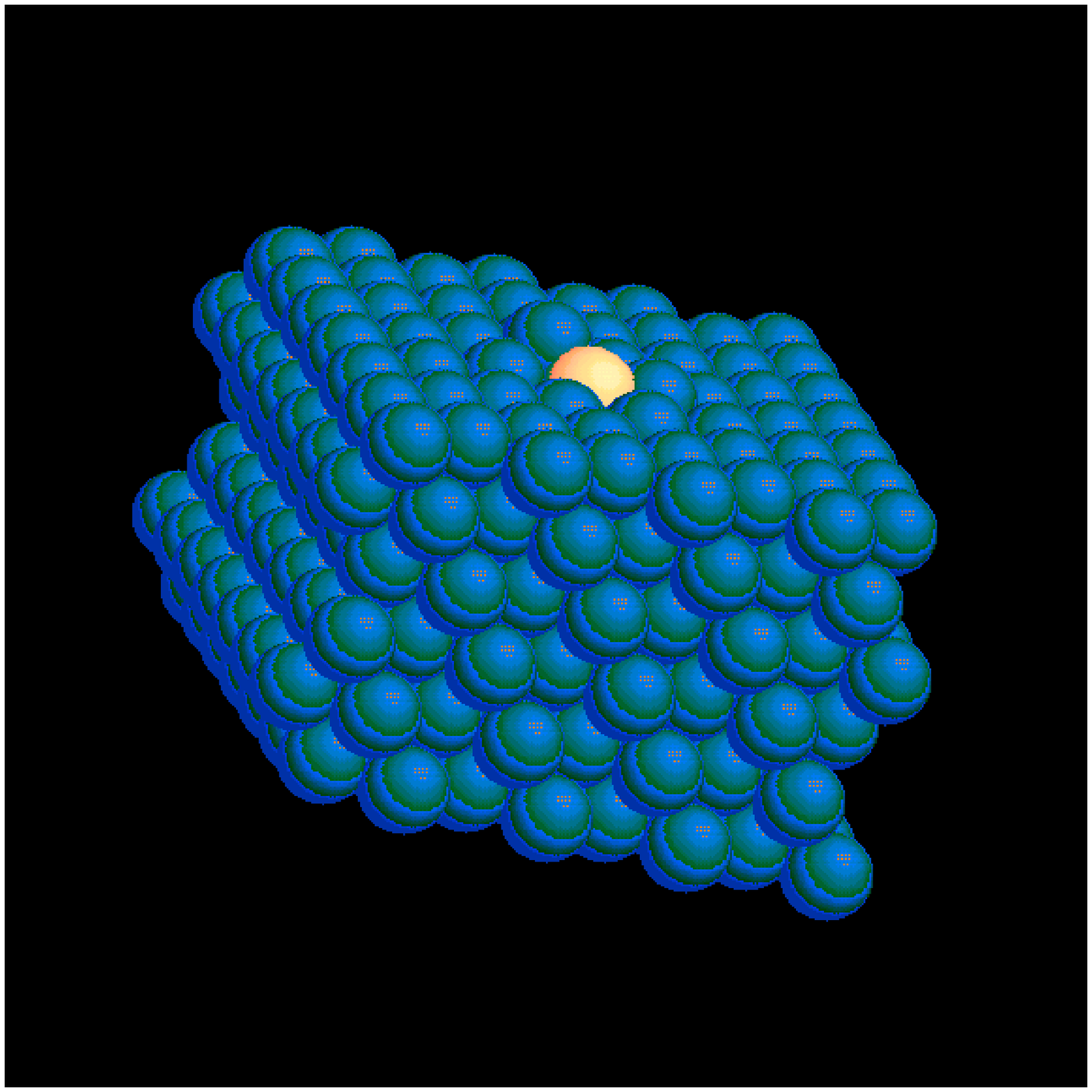}
\includegraphics*[height=3.2cm,width=4.2cm,angle=0.]{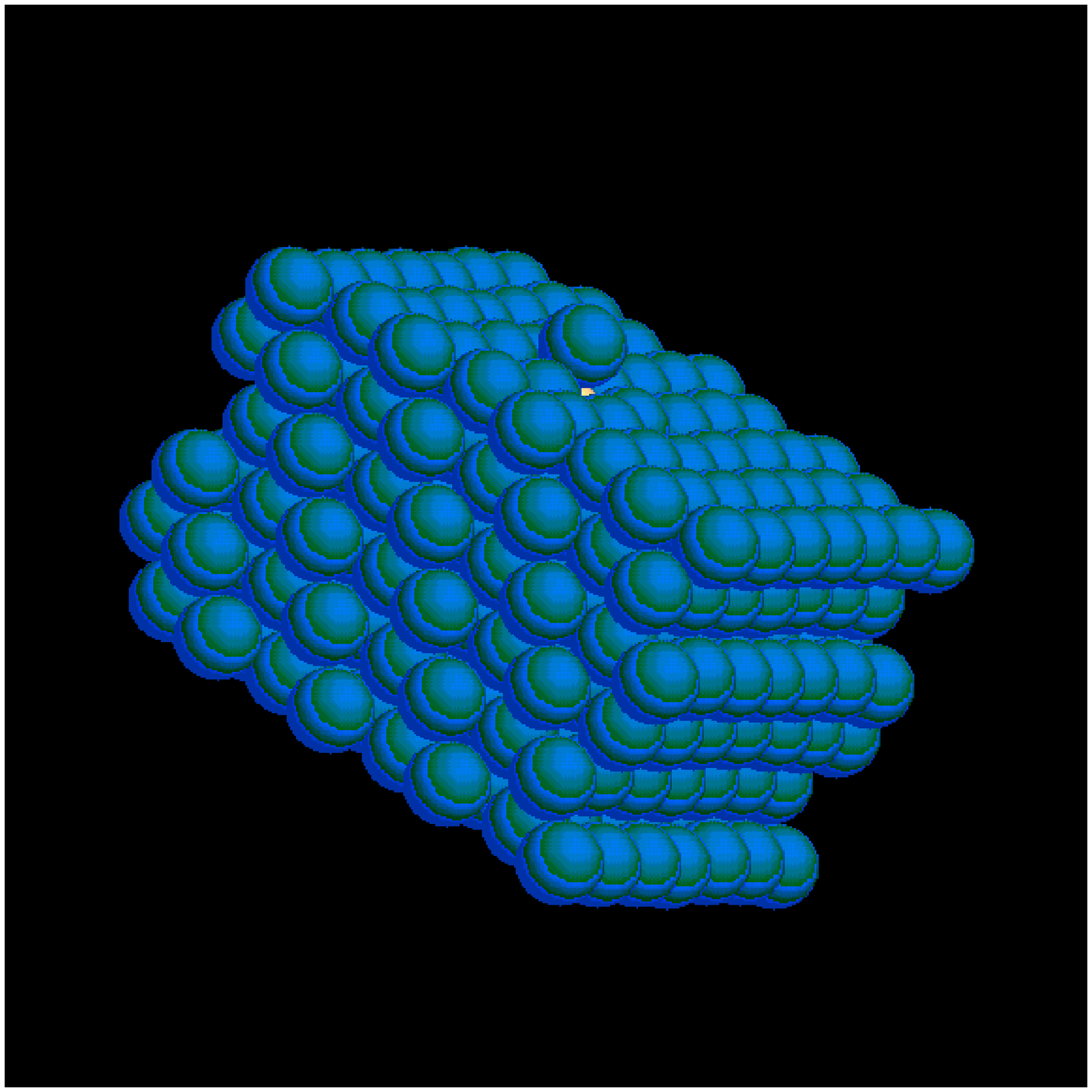}
\caption[]{
The snapshots of the ultrafast atomic injection at $0$, $0.45$, $0.55$, and at $2$, ps (from
left to right)
 $0.8$, $1.2$ and at $1.5$, ps (from left to right)
Pt atom and Al atoms are shown with light and dark (blue) colors,
respectively.
Only few hundreds of atoms of the simulation cell in the "active" region are shown.
The deposited particle is in rest at $t=0$ K (no initial kinetic energy is given).
Therefore the atomic injection occurs spontaneously.
}
\label{figmovie}
\end{center}
\end{figure}

 We also calculate 
 the binding energy $U_b^i$ of the impurity particle $i$ which can be expressed
in terms of its potential energy $U(\bf{r})=$ (summed over interactions
with its neighbors cut off at $r_c$)
and its first derivative (Newtonian forces).
Hence at each time step $U_b^i$ can be calculated from the Newtonian forces.
\beqn
U_b^i=-\sum_{j,r_{ij}<r_c, i \ne j} \int_0^{\infty} \frac{\partial U(\bf{r})}
{\partial
\bf{r}} \biggm|_{r=r_{ij}} \frac{\overline{r_{ij}}}{r_{ij}} d \bf{r}.
\label{binding}
\eeqn
Since the system is energy conservative, the space integral over the Newtonian interatomic forces $\frac{\partial U(\bf{r})}{\partial
\bf{r}}\biggm|_{r=r_{ij}}$ give the total energy of the simulation cell. $\bf{r}$ can be replaced by the internuclear separation $r_{ij}$ in the pair interaction term $U(r_{ij})$.

\section{Results}

{\em Transient inter-layer atomic mixing (TILAM):}
  The simulation of vapor deposition in the Pt/Al(111) system leads to an unexpected result. 
The deposited atoms, independently of the energy of deposition, intermix spontaneously with
ultrafast atomic exchange entering the top Al(111) layer even at $\sim 0$ K within a ps 
leading to the
extremely large jumping rate of $\Gamma \approx 10^{12}$ Hz.
Such a robust rate at $\sim 0$ K has never been reported before in metal system.
Moreover we find that the deposited Pt atom undergoes an abrupt inter-layer
migration spontaneously regardless to the impact energy and to the
strength of the Pt-Al interaction potential.
Hence we find it important to understand the details of this porcess.
Although the obtained rate is surprisingly high,
the employed simulation approach is highly standard and hopefully there is no reason
to question the validity of the results.
\begin{figure}[hbtp]
\begin{center}
\includegraphics*[height=5.5cm,width=7.5cm,angle=0.]{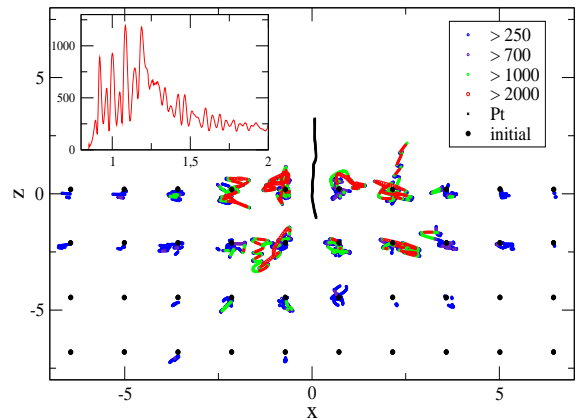}
\caption[]{
The vertical positions of moving atoms as a crossectional view of typical trajectories of
Al atoms in the upper layers (in a crossectional slab cut the middle of the simulation cel
l) induced
by the deposition of a Pt atom at $\sim 0$ K (the scale
is in $\hbox{\AA}$ in the axes, the depth position of the surface is at $z=0$).
The trajectory of the Pt is also shown with a black curve.
The positions of the atoms are collected up to 2 ps during a deposition event.
The different colors of the points correspond to the local temperature
range (K) of the Al atoms in the thermalized region shown above.
{\em Inset on the top:}
The fluctuating local average temperature ($T_{local}(t)$, K) of the thermalized region ($\sim
10 \times 10 \times 5$ $\hbox{\AA}^3$) which includes $30-40$ hot atoms
as a function of the time (ps).
}
\label{traject}
\end{center}
\end{figure}
In particular, we do not think that the result is the artifact of
the employed semiempirical interaction potential.
Our DFT dimer
calculations support 
the reliability of our interpolated
semiempirical potential.
We find such a peculiar behavior only for few diffusion couples (Pt/Cu, Pt/Al,
Au/Al) among those couples for which interpolated CR potential has been available.
Many experimental results support indirectly our finding.
Strong exothermic solid state reactions have long been known between various metals and Al \cite{Buchanan,Barna,Waal}.

  The injection of a Pt atom leads to the ejection of an Al atom to the surface.
The deposition of 1 ML of Pt leads to the formation of an adlayer rich in Al
in agreement
with the experimental findings
\cite{Buchanan,Barna}.
The available room temperature experimental results also report us
strong intermixing for Pt/Al \cite{Buchanan,Barna}.
The computer animation of the atomic injection can be seen in a web page \cite{web}.
We find direct injection only in the case of certain transition metal elements around Pt in the
 periodic table,
such as Ir, Au and also for the Pt/Cu couple.

 {\em The spontaneous local thermalization of the substrate:}
 In order to get more insight into the details of the atomistic mechanism of TILAM we follow
the atomic trajectories of surface Al atoms.
 The TILAM induced surface disordering of Al(111) is shown
in Fig. ~\ref{traject} where the trajectories of the transient vertical
jump of
few Al atoms to the surface (adlayer) can be seen.
We plot the atomic positions of a crossectional slab cut in the middle of the simulation cell
(with a slab thickness of $15$ $\hbox{\AA}$) for the top layer atoms during vapor deposition
of Pt (Fig ~\ref{traject})
at $\sim 0$ K.
\begin{figure}[hbtp]
\begin{center}
\includegraphics*[height=5.7cm,width=6.5cm,angle=0.]{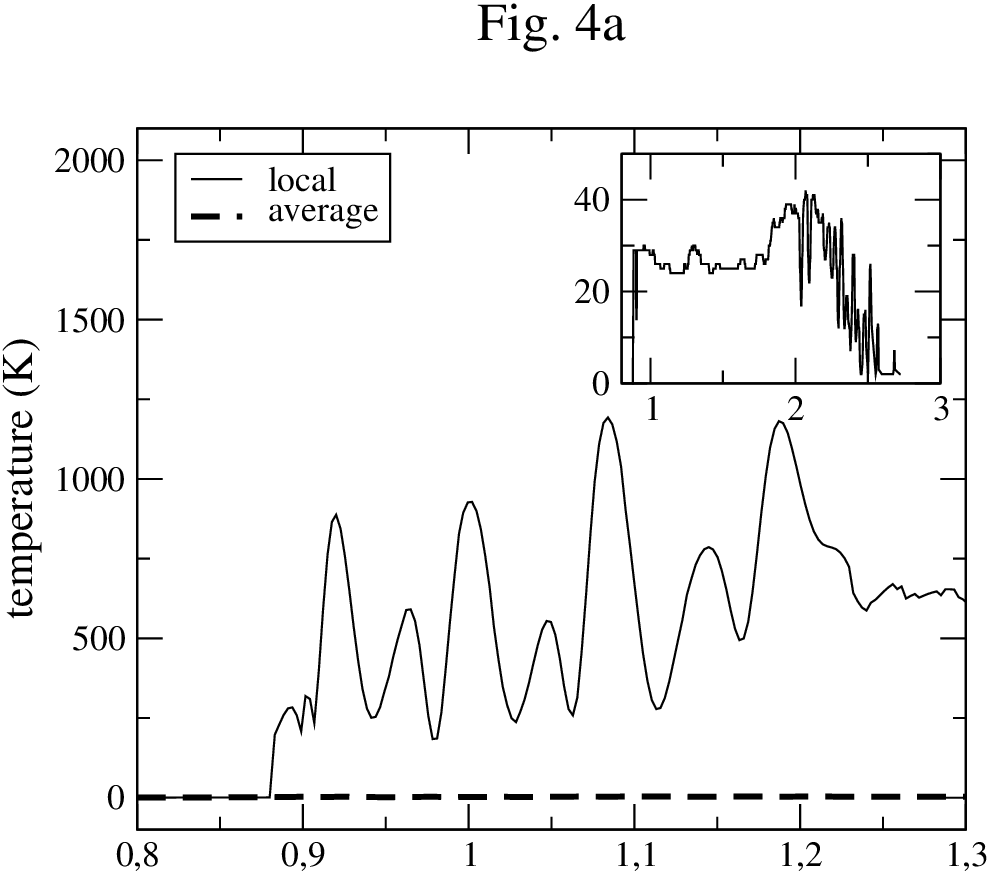}
\includegraphics*[height=5.7cm,width=6.5cm,angle=0.]{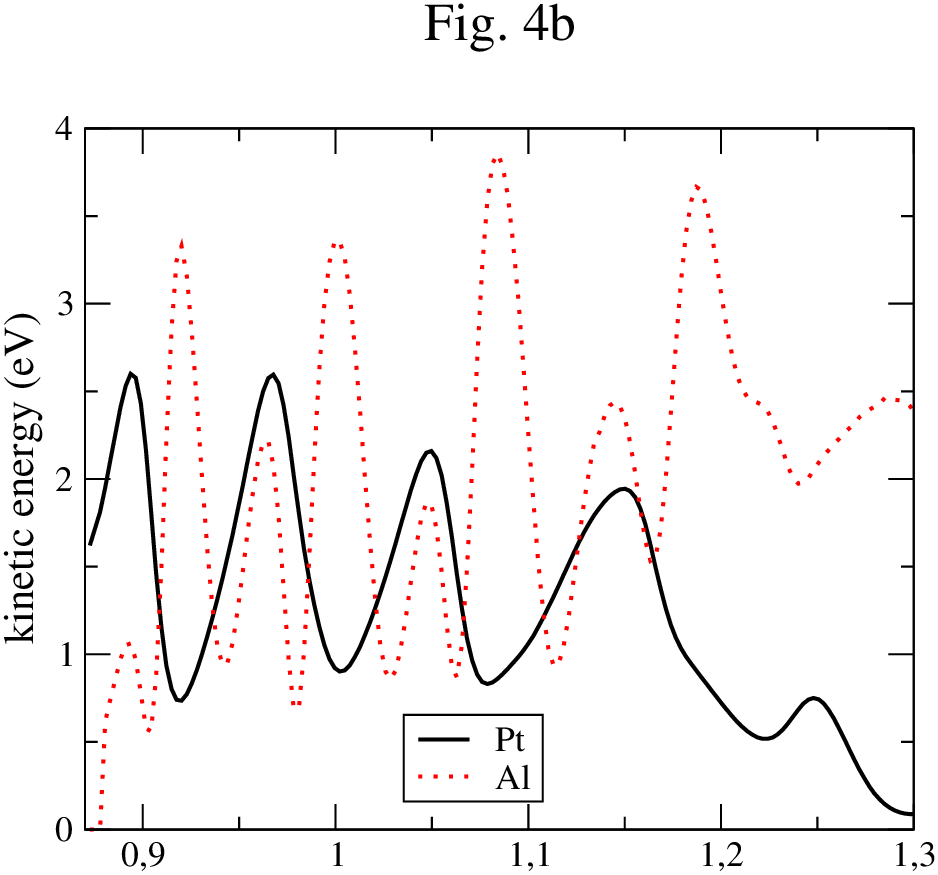}
\includegraphics*[height=5.7cm,width=6.5cm,angle=0.]{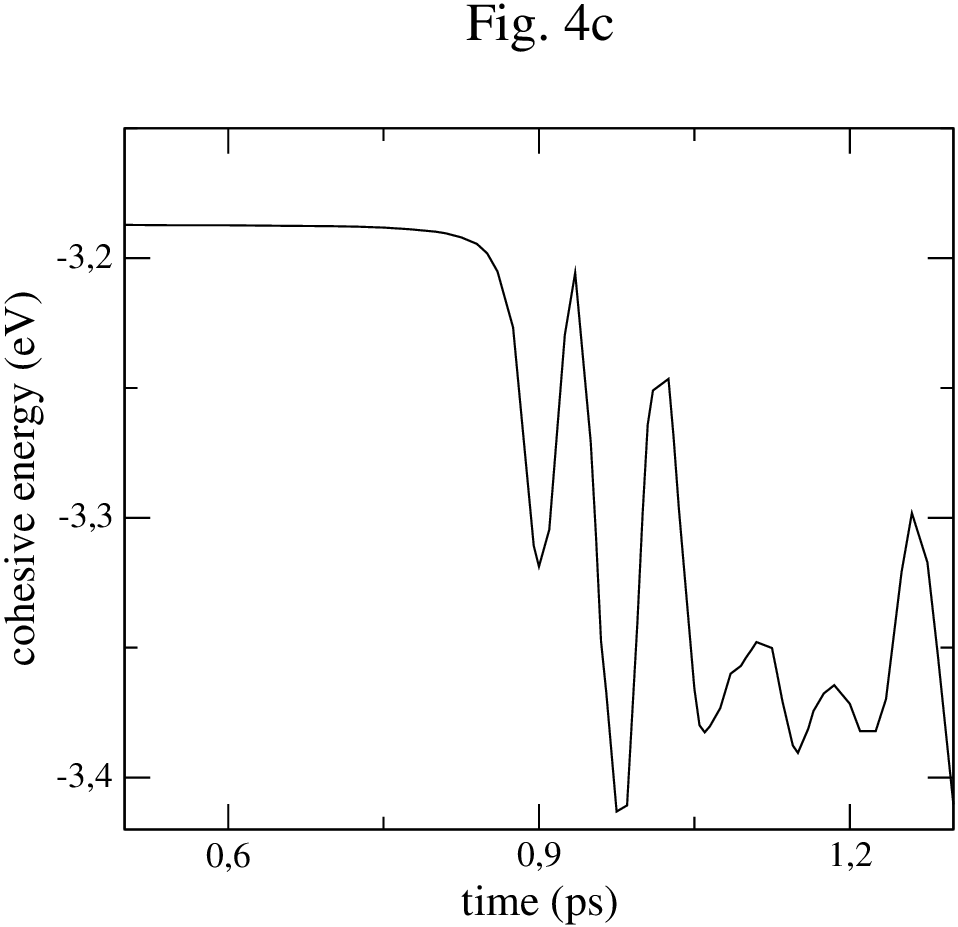}
\caption[]{
{\em Fig 4a:} The local temperature $T_{local}(t)$ (K) of the thermalized region below the surface
(with the volume of $\sim 10 \times 10 \times 5$ $\hbox{\AA}^3$) as a function of
time (ps) obtained for a typical event.
The average temperature of the simulation cell is also shown with
a dashed line.
{\em Inset:} The number of Al atoms in this subsurface nanoscale region as a function of
time (ps).
{\em Fig 4b:} The oscillating kinetic energy (eV) of the impinging Pt atom and of the
transient Al atoms (summed up for the thermalized ensemble) as a function of the time (ps)
shown with a continous and dashed lines, respectively.
The Pt atom has been initialized $4.6$ $\hbox{\AA}$ above the
Al(111) surface with zero velocity.
The initial temperature of the Al cell is nearly zero.
{\em Fig 4c:} The average cohesive energy/atom (eV) in the thermalized region of Al
as a function of time (ps).
}
\label{kinetic_Pt_Al}
\end{center}
\end{figure}
The mobility of the substrate atoms is large with large amplitudes
around their equilibrium positions.
This is surprising because
the overall temperature of the entire cell does not exceed few K during the simulations.
However, we find that within a small volume below the surface including few tens of atoms the
local temperature can be surprisingly high.

  \subsection{Local temperature within a thermalized subsurface nanoscale zone}

An approach will be outlined briefly which has been used for obtaining continuously distributed local properties from the positions and velocities of constituent atoms
obtained by MD simulations.
The applied methodology is similar to that obtained for analyzing the results
of MD simulations for systems with finite size and which can not be described
by continuum models \cite{localtemp,Ikeda}.
Unfortunatelly, when the system size is shrinked to the nanoscale fluctuations, such as the
spatial oscillation of the temperature will be enhanced \cite{Ikeda}.
Therefore specific definitions are needed for giving
time dependent local quantities, such as the local temperature of a
nanoscale system.

 Thermondynamic quantity, such as the temperature $T$ can only be assigned
to an atomic ensemble in which the number of particles $N$ is sufficiently
high to exclude the effect of local fluctuations (statistical ensemble average).
Unfortunatelly, a nanosystem, which often includes less than $\sim 1000$ atoms
(nanoclusters) can not be described by $T$ in a conservative point of view.
In these cases, however, one should introduce the quantity local temperature $T_{local}$
which can be used to explain the thermal properties of nanostructures.
We define $\langle T_{local} \rangle$ as the time averaged temperature of the thermalized sub(nano)region
obtained during simulations sampling the sufficiently large portion of the phase space \cite{Frenkel}.
As a natural consequence, $\langle T_{local} \rangle \rightarrow T$, when $N \rightarrow \infty$,
or $N$ is sufficiently large to have a statistical meaning of the ensemble.

  The total simulation cell of Pt/Al during the intermixing of Pt is a highly anisotropic and inhomogeneous system.
In this case the thermalized region of the substrate
can be taken as a nanosystem
including few tens of hyperthermal atoms at the surface,
however, which is not isolated from its low-temperature environment.
There is a continous thermal exchange between the "hot spot"
and the ultra-low temperature environment and with the heat bath.

 Much effort has been put forward in establishing a relationship
between thermodynamic quantities and MD data \cite{Allen,Frenkel,Ikeda,Jellinek,Berry}.
The {\em ergoditic theorem} ensures in thermodynamics the relation between
the observable ensemble averages and the simulated time averaged quantities \cite{Allen,Frenkel}.
Under ergodic condition the time average or ensemble average of the velocity
distribution of the constituents will closely follow the Maxwell-Boltzmann
distribution.
MD studies provides microscopic (atomic) information and an effective
temperature can be derived from the individual atomic velocities \cite{Allen,Ikeda}.
Another useful quantity,
the kinetic temperature of a nanosystem is defined as an averaged kinetic energy
per an individual spatial degree of freedom \cite{Jellinek,Berry}.
   The time averaged temperature of the nanosystem with $N$ number of atoms can be given as
\beqn
 \langle T_{local} \rangle^N =  \lim_{t \rightarrow \infty} \frac{1}{t} \int_{t=0}^{\infty} T_{local}(t) dt ~~~~~~~~~~~ && \nonumber \\ = \lim_{t \rightarrow \infty} \frac{1}{t} \int_{t=0}^{\infty}
\frac{2 E_{kin}(t)}{3 k_B} dt && \nonumber \\  
~~~~~~~~~~~~~~~~~~\approx \frac{1}{M} \frac{1}{N} \sum_j^M \sum_i^N \biggm\{ \frac{2 E_{kin}^i(t)}{3 k_B} \biggm\}_j,
\eeqn
where $M$ is the number of visited configurations in the simulations.
The instanteneous value of $T_{local}(t)$ will fluctuate around the mean value $\langle T_{local} \rangle^N$ unless an infinite number
of particles haven been considered.
Indeed, under the assumption of ergodicity, this mean value exists and is independent of the initial data with identical energy.
In nanosystems ergodicity looses its validity and we can not give
any macroscopic observable which is related to $\langle T_{local} \rangle$.
Nevertheless, the time evolution of $T_{local}(t)$ could be a useful quantity
to monitor system changes during transient phase evolutions such as
e.g. local melting transition \cite{Sule_NIMB04}.

  The calculation of $T_{local}(t)$ within an arbitrarily small volume
including $N$ particles can also be calculated formally, however, from the
kinetic energy of the individual particles obtained from simulations in each time step.
The {\em equipartition theorem} (ET) allows in principle to relate the temperature
of a system with its average energy.
If the mobility of atoms is sufficiently high within the thermalized nanoregion, e.g. the number of
hyperthermal atoms is large, we are close to the limit of
classical ideal gas. In this case 
each of the mobile particles has an average kinetic energy of $(3/2)k_B T$ in thermal equilibrium, where $k_B$ is the Boltzmann constant and $T$ is the temperature. 
However, our system should not be strictly an ideal gas.
ET requires only that $k \langle T_{local} \rangle \gg h \nu$, where
$\nu$ is the phonon frequency of an oscillator in the thermalized zone, hence
in highly excited
states quantum effects should become negligible \cite{Born}.

 Although our system is not in a thermal equilibrium, we can expect that
the ET works also nearly correctly for those cases which are not very
far from equilibrium.
It is also known that the spatial-temporal variation of $T_{local}(t,r)$ can be calculated
for even highly inhomogeneous systems such as e.g. a plasma, shock loaded systems \cite{localtemp}
or for ion-bombardment induced thermal spikes \cite{Sule_NIMB04}.
The thermalized region if assumed as an overheated liquid state of matter,
and is close to an ideal gas and than ET can be applied.
In an ideal gas atoms can move few $\hbox{\AA}$/ps (that is a nearly ballistic atomic mobility).
In our system we find a similar rate of mobility within the thermalized region during simulations
when a Pt impurity atom has been injected.
The thermalized state of the local subregion in Al persists up to few ps
which is a very short lifetime in the thermodynamic sense.
The quench rate of the liquid-like phase is extremely fast.
Nevertheless,
the local temperature $T_{i,local}(t)$ can be formally assigned to each particles 
during the molten phase
at time $t$ using the ET,
\be
\frac{1}{2} m_i v_i^2(t) = \frac{3}{2} k_B T_{i,local}(t),
\ee
where $v_i(t)$ and $m_i$ are the atomic velocity and mass of $i$th
particle. 
Note that we do not follow the spatial variation of $T_{i,local}(t)$ as it has
been calculated e.g. in ref. \cite{localtemp}.
$T_{i,local}(t)$ is always assigned to the velocity $v_i(t)$ of particle $i$ at time $t$.
We are interested in the time evolution of $T_{i,local}(t)$ in a particle trajectory.
$T_{i,local}(t)$ has no physical meaning in a strict thermodynamic sense since ergodic
theory relates only the time averaged $\langle T_{local} \rangle$ to the
ensemble average (observable temperature).
However, summing up for each particles within the thermalized volume and calculating the average $T_{local}(t)$ of an ensemble of particles in a local volume
could provide a time dependent ensemble averaged local temperature with physical meaning.
The averaged local temperature $T_{local}(t)$ within a region of the substrate is given
than at time $t$ for $N$ number of
hyperthermal atoms
\be
T_{local}(t)= \frac{1}{N} \sum_i^N T_{i,local}(t)
=\frac{1}{N} \sum_i^N \frac{m_i v_i^2(t)}{3 k_B}.
\label{T}
\ee
We use $T_{local}(t)$ for describing the time evolution of local heating up (thermalization) processes.
Probing few tens of configurations with different starting positions for Pt
we get very similar events hence we do not average for events in 
Eq. ~\ref{T}.
Hence the plotted $T_{local}(t)$ values do not correspond to
an ensemble average, instead we show the time evolution of
the simulated $T_{local}(t)$ for a typical Pt/Al deposition event 
in 
Fig. ~\ref{kinetic_Pt_Al}a.
Therefore the plotted $T_{local}(t)$ curve is not a unique one,
the shape of temperature fluctuation changes event by event.
However, in most of the events we find similar features:
multiply peaked structure (transients) and high average
local temperature within the subsurface zone $\langle T_{local} \rangle \approx 500$ K during the deposition.
Such kind of an analysis can be used e.g. for studying the time evolution of local melting in collisional cascades and thermal spikes \cite{Sule_PRB05,Sule_NIMB04}.

 The magnitude of $T_{local}(t)$ clearly depends on the volume of the subregion considered
in the summation in Eq. \ref{T}.
We choose a subregion in which the individual kinetic energy of particles
exceeds a threshold value (that is $E_{kin} \ge \frac{3}{2} k T_{melt}$, where $T_{melt}$
 is the bulk melting point of e.g. Al,  $E^{Al}_{kin} \approx 0.07$ eV/atom in this case).
This threshold value is rationalized by our experience reported in recent
publications \cite{Sule_NIMB04}
in which we found that local melting transition can be described by
the occurrence of sufficiently large number of "liquid" particles which possess 
$E_{kin} \ge  \frac{3}{2} k T_{melt}$.
Out of the subregion we cut off contributions to Eq. ~\ref{T}.
The crossection of the thermalized volume at and below the surface
can be seen in Fig. ~\ref{traject} which shows us that the volume of this region
is less than $\sim 10 \times 10 \times 5$ $\hbox{\AA}^3$ including less than $\sim 200$ atoms.

  Using the local temperature analysis outlined above we find that in the core of the disordered region ($20-40$ atoms in the subsurface zone of
$\sim 5 \times 5 \times 2.5$ $\hbox{\AA}^3$)
the peak local temperature (LT) 
reaches $T_{local}(t) \sim 1000$ K for few times during the persistense of thermalization (see Fig. ~\ref{kinetic_Pt_Al}a).
In the large zone of $\sim 10 \times 10 \times 5$ $\hbox{\AA}^3$ 
$T_{local}(t) \sim 100-150$ K.
This is really surprising that $T_{local}$ can reach so high instanteneous values while the
external temperature is $\sim 0$ K. This is because no
external forced condition has been applied:
simply the presence of an impurity particle induces the
local thermalization.
This can be taken as a spontaneous process: no driven conditions has been applied.
The impurity particle has been added to the host system in a rest since no
initial velocity has been given to it.
We are not aware of other reports in which the spontaneous local heating up
of a $\sim 0$ K system has been published before.
This finding could explain the strong exotermicity known during
alloying between various transition metals and Al at room temperature 
which could even lead to extremely fast burn rates \cite{shock}.

 \subsection{The driving force of TILAM: kinetic energy transfers}

 We also find that the system exhibits a complicated time evolution.
We plot various local properties against time, such as
$T_{local}(t)$, the number of Al atoms in the thermalized subregion
(Fig. ~\ref{kinetic_Pt_Al}a),
the kinetic energy of the Pt atom and the Al atoms (the sum of the kinetic energy of
the thermalized Al atoms, Fig. ~\ref{kinetic_Pt_Al}b) and
the average cohesive energy in the thermalized nanoregion (Fig. ~\ref{kinetic_Pt_Al}c).

 We notice in Figs. ~\ref{kinetic_Pt_Al} the multiply peaked structure of the curves. 
The kinetic energy of Pt oscillates as a function of time and which correlates
with the fluctuation of the LT of the thermalized region of Al shown
in Fig. ~\ref{kinetic_Pt_Al}a.
The correlation also holds with
the average kinetic energy of the hot Al atoms
shown in Fig. ~\ref{kinetic_Pt_Al}b).
In particular, we see 4 nearly equidistant peaks separated by $\sim 0.1$ ps
in Fig. ~\ref{kinetic_Pt_Al}b.
This peculiar feature of the system does not support a heat spike mechanism
(that is the simple impact induced thermalization and local melting).
The low impact energy of $\sim 2$ eV alone (obtained during local acceleration) is insufficient for the
appearance of a collisional cascade not even for a thermal spike
(the formation energy of a vacancy-displaced atom pair, Frenkel pair is $\sim 25$ eV).
There must be an additional mechanism
which thermalizes the top region of the Al substrate.

 We would like to emphasize the spontaneous nature of the local heating up
because we let the system to time evolve during the simulations without
\begin{figure}[hbtp]
\begin{center}
\includegraphics*[height=5.9cm,width=7cm,angle=0.]{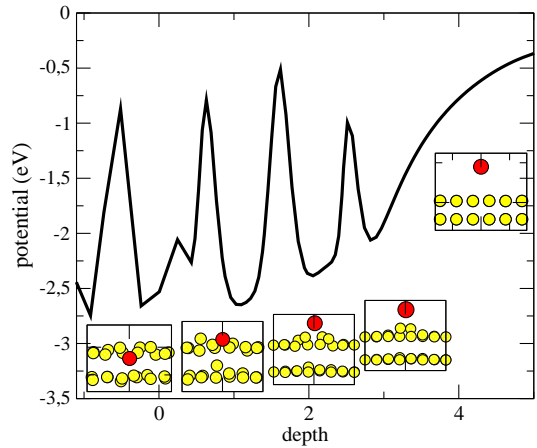}
\caption[]{
The dynamical potential energy (($U(z)$), binding energy) of the impinging Pt atom (eV) during a single atomic jump through the topmost layer of Al(111) as a function
of the distance from the surface (in $\hbox{\AA}$).
The crossectional views of the thermalized region of the system are also shown
with the impinging impurity particle
for the initial system and for the transition (activated) structures (which correspond to the
barriers) at $t=0$, $0.85$, $0.95$, $1.05$ and $1.15$ ps.
}
\label{oscillatory}
\end{center}
\end{figure}
any external perturbation.
No kinetic energy is given to the impurity atom, hence no
externally forced condition has been applied.
However, the system reorganizes itself spontanenously.
This is typically the characteristics of a self-organizing system.

\subsection{The dynamic adsorbate-surface potential energy profile:}

  In Fig. \ref{oscillatory} we show the dynamic potential energy profile of the approaching 
Pt atom together with the
crossectional views of the structures corresponding to the energy barriers
of the oscillatory potential.
The potential energy of Pt as a function of the distance from the surface
($U(z)$) has been calculated using Eq. ~\ref{binding}.
The barrier height of these peaks is around few eV.

 The static Pt-Al(111) potential energy profile (and surface), which could be calculated e.g. by
{\em ab initio} density functional approaches, does not account for the dynamic
nature of TILAM. We see from the animations \cite{web}
the strong mobility of the substrate atoms during the ultrafast atomic exchange.
Hence the dynamic Pt-Al(111) potential energy profile, obtained by MD, describes more correctly
the energetics of the system.

 In Fig ~\ref{oscillatory} we show the binding energy (potential energy) of the Pt atom as a function of the
distance from the surface.
The binding energy has been calculated as a sum of pair interactions between the Pt and Al atoms
up to the second neighbors.
Further neighbor interactions (3rd and higher orders) have vanisingly small contribution.
For typical events multiply peaked
\begin{figure}[hbtp]
\begin{center}
\includegraphics*[height=5.5cm,width=7cm,angle=0.]{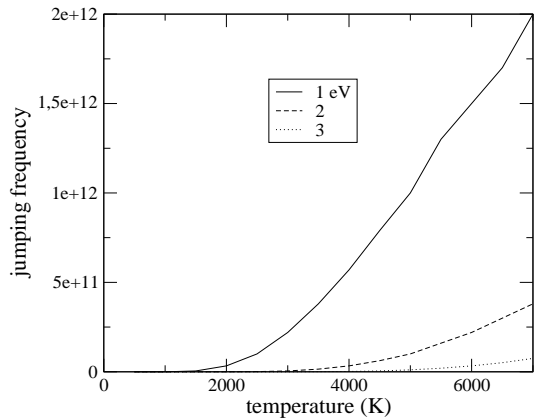}
\caption[]{
The thermally activated jumping frequency (m$^2$/s) as a function of $T_{local}$
obtained by using the
expression of $\Gamma = \Gamma_0 exp (-\frac{E_a}{k T})$ at different
activation energies given in eV.
}
\label{Dth}
\end{center}
\end{figure}
oscillatory impurity-substrate interaction potential have been found.
It must be emphasized that
the profile of the potential is not unique, e.g.
the depth positions and the height of the barriers slightly scatter in various events, nevertheless, the
oscillatory behavior typically occurs for all of the events during which
injection take place.

The magnitude of the scatter does not exceed few tenths of eV for the barrier height and
few tenths of $\hbox{\AA}$ for the depth positions of them.
Due to the nonuniqueness of the oscillatory behavior
it is no use to plot a profile over a statistical average of various events.
It must also be noted that a potential energy profile for a normal deposition event without
injection (when the impurity atom becomes an adatom) is a unique single well potential.

 Oscillatory interaction profiles have already been reported on the surface of metals
between
adatoms \cite{oscill}, however, no reports have been found for IM.
The deposited particle could go through the multiply barriered potential
due to the kinetic energy obtained during the local acceleration towards the surface
and due to a still unknown mechanism which will be characterized in the next section.
The magnitude of the peak kinetic energy (arrival energy) is in the range of few eVs.

 The particle can gain the adsorption energy (condensation energy)
as a kinetic energy during the accomodation process on the surface.
Sufficiently high kinetic energy gain allows the transient lateral movement of the particle
on the surface until sticking (trapping) occurs \cite{Michely}.
In Pt/Al, however, we find that instead of the lateral transient mobility the
deposition of the impurity particle leads to transient IL atomic mobility.

\section{Discussion}

 \subsection{Quantum tunneling diffusion is ruled out:}

 Since low-temperature and ultrafast atomic transport has been known as quantum tunneling
it could also be that the impurity injection of Pt to Al(111) is also QD.
In principle, classical molecular dynamics can account for QD implicitly via the adjusted the parameters,
although no direct quantum effects has been incorporated into the interaction potentials.
E.g. the QD of H in various metals can be simulated by classical MD.
\cite{QD_H}.
Therefore we check this process for QD.
The de Broglie wavelength of a Pt atom $\lambda=2 \pi \hbar/ \sqrt{3 m k T} \approx 1.7$
$\hbox{\AA}$ (during the simulations we find $T \approx 3$ K in the simulation cell) which is somewhat less than the
inter-layer hopping distance
of $\sim 2-3$ $\hbox{\AA}$.
However, if we estimate the tunneling jump rate $\Gamma$, which is approximated by
\cite{Bulou2,Merzbacher}
\be
\Gamma_{tun} = \frac{2 \omega}{\pi^{2/3}} \sqrt{\frac{2 E_a}{\hbar \omega}}
exp \biggm(-\frac{2 E_a}{\hbar \omega} \biggm),
\ee
where $\omega=\sqrt{2 E_a/m . b^2}$, $b$ is the barrier width ($b \approx 1$ $\hbox{\AA}$).
We get a vanisingly small $\Gamma_{tun} \approx 10^{-22}$ Hz, which is very far from our finding of $\Gamma_{MD}
\approx 10^{12}-10^{13}$ Hz ($\sim 1$ inter-layer jump/ps).
We conclude that atomic transport with quantum tunneling can not explain the occurrence of transient inter-layer atomic mobility and
there must be a peculiar mechanism which promotes impurity particle acceleration
through the top layer of Al at $\sim 0$ K.
Using the Van't Hoff-Arrhenius expression (which can be used for classical atomic transport) of $\Gamma = \Gamma_0 exp ( -\frac{E_a}{k T}
)$ gives also zero thermal rate of $\Gamma \approx 0$ Hz
(the preexponential $\Gamma_0 \approx 10^{13}$ Hz \cite{Philibert}).

 The dynamic activation energy is taken from Fig. 4, from the dynamic potential energy profile of Pt along the reaction coordinate ($E_a=1-2$ eV).       
However, if we take into account the local heating up ($T_{local} \approx 1000$ K of the region where
the impinging Pt atom is intermixed) we get
$\Gamma \approx 10^{3}-10^{8}$ Hz which is also far from our MD simulations.
Although in the hot core of the thermalized region we find few thousands of K temperature,
however this does not allow such an increase of the thermal jumping frequency.
This has been demonstrated in Fig. ~\ref{Dth}, in which we plot
the thermally activated $\Gamma(T_{local})$ as a function of the local temperature in
the thermalized nanoregion. We show the obtained curves with $3$
different activation energy values.
It can be seen that even at $1$ eV activation energy, which is below our
calculated value ($E_a \approx 2-3$ eV) , we get $\Gamma$ much below the simulated
$\Gamma_{MD}
\approx 10^{12}-10^{13}$ Hz in the reasonable temperature regime
($T < 2000$ K).

 The rate can only be in accordance with the simulated value within
the Arrhenius picture
if the prefactor $\Gamma_0$ is increased by few orders of magnitude
or at very high temperature of $T \approx 10^{5}$ K.
However, recent results rule out the occurrence of anomalous preexponential
factors \cite{Michely}.
\begin{figure}[hbtp]
\begin{center}
\includegraphics*[height=5.5cm,width=7cm,angle=0.]{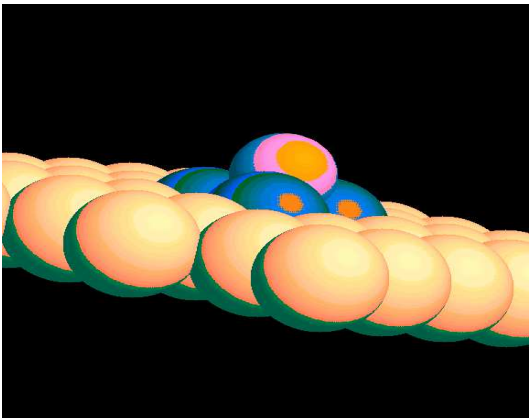}
\caption[]{
The snapshot of the transient facilitated intermixing (also called as TILAM) process driven by the assistance of few surface Al atoms (which move out-of-plane coherently) at $0.95$ ps.
The displaced Al atoms are shown with a darker color.
The impinging impurity particle (Pt) is shown with a lighter color on top of
the displaced Al atoms.
Only the central region of the top layer is shown.

}
\label{triad}
\end{center}
\end{figure}
The serious deviation of the jump rate from the Van't Hoff-Arrhenius equation
is highly unusual \cite{Philibert} and has only been found for such cases
when $E_a \approx k T$ \cite{Michely} which often lead to superdiffusion.
The discrepancy between the thermal and transient rates could be accounted for
assuming an auxiliary mechanism which facilitates the amplification of
intermixing.
In the rest of the paper we outline the details of such a possible mechanism.

 \subsection{The kinetic energy oscillation of Pt:}

The downward momentum transfer provided by the transient out-of-plane Al atoms on the surface
surpasses the short-ranged impurity-host repulsion which promotes
TILAM through an energeticaly unfavorable transtion state.
The coincidence in time of peaks of the time evolution of
the kinetic energy of the impinging Pt atom in Fig. ~\ref{kinetic_Pt_Al}b and
the local temperature $T_{local}(t)$ in Fig. ~\ref{kinetic_Pt_Al}a
in the thermalized region supports this explanation.
Moreover, we see in Fig ~\ref{kinetic_Pt_Al}b
that the sum of the kinetic energy of the transient Al atoms
reaches its maximum with some time delay when compared with
the kinetic energy of the Pt atom Fig. ~\ref{kinetic_Pt_Al}b. The delay occurs 4 times
indicating that the momentum transfer really occurs.

 First the approaching Pt particle gains kinetic energy during surface
acceleration and gives the obtained energy to the Al surface
which in turn is heated up locally (a peak occurs with some
time delay for the Al atoms).
Than a second lower peak appears for the Al atoms together with
a higher peak for the Pt atom which supports the idea of momentum
transfer between the transient Al atoms and the Pt atom.
This process is repeated for few times periodically.

\begin{figure}[hbtp]
\begin{center}
\includegraphics*[height=5.5cm,width=7cm,angle=0.]{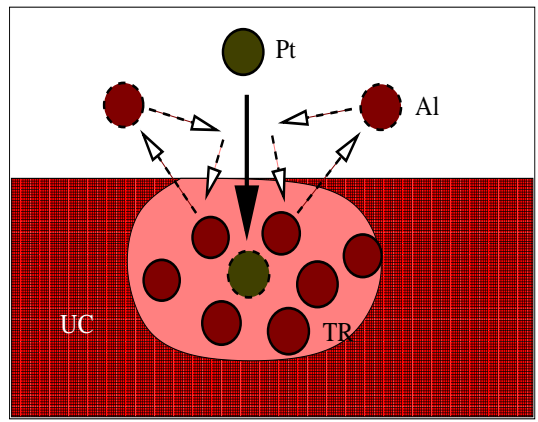}
\caption[]{
The schematic view of transient facilitated intermixing in the
coexisting solid-liquid like system.
TR and UC denote the thermalized subsurface nanoregion ($T_{local} \approx 1000$ K) and
the ultracold substrate (few K).
Opened arrows with dashed line show
the trajectory of out-of-plane Al atoms which give
downward momentum to the approaching Pt atom pushing down it
below the top layer of Al(111).
}
\label{schem}
\end{center}
\end{figure}
 In Fig ~\ref{triad} we show the appearance of few transient Al atoms on the surface
at $t \approx 0.95$ ps which catalyse the injection of Pt.
In particular, $4$ surface atoms move out-of-plane coherently and opens up a channel
for TILAM.
The average distance of these atoms from the surface is not larger than $\sim 1$ $\hbox{\AA}$.
At a critical proximity of the Pt atom to the surface ($d_{PtAl} \le 3$ $\hbox{\AA}$)
strong, mostly out-of-plane atomic displacements of few Al atoms at the surface
set in
which leads to the injection of the particle.
Without the out-of-plane surface instability no injection occurs, the particle becomes an adatom.
Hence the {\em concerted motion} of few surface atoms in a self-organized manner opens up a channel spontaneously for injection.
The temporally ejected and returning transient Al adatoms give a downwards momentum to the approaching 
impurity particle: this kind of a kinetic energy transfer is so effective that               the impurity particle can move through the top layer although the                            dynamic barrier height of this process is few eV.       

 \subsection{The microscopic mechanism of TILAM:}

  The obtained results in this article might support the following mechanism:
The impurity Pt atom approaches the surface of Al(111) by surface acceleration.
The deposited particle is injected to the substrate by few out-of-plane
vibrations of few surface Al atoms in 4 or more steps. 
The Pt atom gains the kinetic energy of the transient
Al atoms which promotes atomic intermixing.
The schematic details of the process can be seen
in Fig. ~\ref{schem}. 
The schem shows us that few transient Al atoms at the surface 
move out-of-plane and when returning back give downward momentum
to the approaching Pt atom injecting it below the surface.

 In Fig. ~\ref{oscillatory} we see, that the energy difference between the initial
and final configurations (between the energy minima) is $\sim -0.5$ eV which is also
the driving force of TILAM (although it is insufficient alone for barrier crossing).
Also the TILAM of Pt is driven not purely by the surface acceleration of Pt:
first the impurity atom accelerates towards the surface, than slows down 
and transfers its kinetic energy to few of the Al atoms (local heating up)
and in the next period gains kinetic energy again provided by the vibrating Al atoms and this is repeated for few times.
Hence the ultra-low temperature IM of Pt is accomplished via a peculiar mechanism
which includes mutual kinetic energy transfers between few substrate atoms
and the impurity atom.
The approaching Pt atom induces local heating up (thermalization) 
and in turn the transient Al atoms provide back kinetic energy to the Pt atom.
This {\em mutual transfer of kinetic energy} takes place few times which
amplifies inter-layer atomic transport substantially.
The spontaneous interplay of kinetic energy transfers between the impurity and
hyperthemal host atoms is a self-organized process:
no external load of energy or perturbation is needed to initialize 
transient facilitated intermixing (TFI).
TFI proceeds via the coherent movement of few atoms in the thermalized nanoscale subsurface zone
induced by the impinging Pt atom.

 \subsection{The possible role of atomic mass anisotropy}

 The importance of mutual kinetic energy transfers during TFI (TILAM) is further evidenced
by the simulation result that if we interchange atomic masses between Pt and Al
(this can be done in MD without changing other parameters) and further
increase the atomic mass of Al to $m_{host} \ge 800$ g/mol
TFI can be suppressed.
Surpisingly this very small atomic mass anisotropy of $\delta < 0.03$,
 ($\delta = m_{imp}/m_{host}$, 
where
$m_{imp}$ and $m_{host}$ are the atomic masses of the impurity and host atoms, respectively)
is required to stop intermixing of Pt while the natural $\delta \approx 7.5$.
Hence in Pt/Al we find that TFI is rather insensitive to the variation of the mass anisotropy.

 However, TFI can also be induced in other impurity/host couples with sufficiently
large atomic mass anisotropy where
$m_{imp} \gg m_{host}$.
Indeed, if we set in artificially large $\delta=6-8$ e.g. in Ni/Al or in Cu/Al which do not show TILAM with natural $\delta$, we also observe
TFI.
No TILAM occurs in other mass anisotropic systems, such as e.g. Au/Ni ($\delta \approx 3.4$) or in
Pt/Ti ($\delta \approx 4$) at natural mass anisotropy.
However, setting in $\delta > 7$ in Pt/Ti, TILAM can also be induced.
In Au/Ni we could not induce TILAM even with extremely large $\delta$.
Nevertheless, we conclude from this that $\delta$ plays some role, however, not only $\delta$ governs TFI
as the prototypical Pt/Al system shows it.
It should also be noted that 
the role of $\delta$ in various intermixing processes
has already been studied in detail recently \cite{Sule_JAP07,Sule_PRB05,Sule_SUCI,Sule_NIMB04}
and strong mass effect has been found during various ion-bombardment induced
intermixing processes.

\subsection{Repulsive intermixing and negative mobility:}

One of the most intriguing consequence of the peculiar mechanism of TILAM is that 
since the process is purely kineticaly driven the strength of the Al-Pt interaction
does not affect IM.
We find that even if repulsive crosspotential is used, TILAM occurs.
Tuning the strength of Al-Pt interaction the speed and the frequency of atomic injection
is unaltered.
Only very strong repulsive potential suppresses TILAM.
 The transient mobility of the impinging particle against the repulsive potential
of the surface can also be understood as a transient negative mobility.
This is similar to that reported by R. Eichhorn et al. for Brownian
motion \cite{ANM}.
Negative mobility of a particle occurs when the response of a system to an external load
(applied bias) is opposite to the direction of this applied force.
In our case if we place an impurity particle (that is Pt atom) in rest above the repulsive
surface of Al(111)
the particle begins to accelerate towards the surface, against the repulsive field
providing a negative response and mobility.

 Hence in the facilitated transport system of Pt/Al(111)
the facilitated atomic transport is largely independent of the
forces of interaction between the intermixing impurity atom and the
substrate (host) surface.
A conventional thermally activated inter-layer transport 
proceeds at the expense of the binding force developed in the transition state.

\subsection{Possible indirect experimental evidences of TILAM}

(i) The occurrence of an Al adlayer upon monolayer Pt deposition \cite{Barna}
which is in accordance with our finding that after the injection of Pt
an Al atoms is released to the surface as an adatom.
\\
(ii) The strong exothermicity \cite{shock} and large negative heat of mixing \cite{Waal} observed in various metal-Al reaction during thin film growth
could be the macroscopic fingerprint of local thermalization observed in the nanoscale.
\\
(iii) The large intermixing length reported recently \cite{Buchanan} also supports
our findings: repeating for many times Pt deposition we also get a strongly intermixed
film. 
\\
(iv) The strong ion-bombardment induced intermixing in Al/Pt bilayer \cite{Gyulai}
is reproduced by our simulations \cite{Sule_SUCI}.
These findings are reproduced at least qualitatively by our Al-Pt potential hence
we are convinced that the heteronuclear potential adequatively describes
interatomic interaction between Al and Pt.

 Finally, the direct experimental confirmation of this new atomic transport mechanism
could be carried out by ultra-low temperature scanning tunneling
microscopy measurements for submonolayer deposition events.

\section{Conclusions}

 We predict an unprecendented rich dynamic phenomenon that has not been 
previously anticipated.
 We found a unique situation in which an impurity particle
can move spontaneously through the barrier without reflection and
external energy income even at $\sim 0$ K.

 The microscopic mechanism of the anomalously fast inter-layer transport 
of the deposited impurity atom
at $\sim 0$ K in few film/substrate couples (X/Al(111), where X=Pt, Au, Ir and in Pt/Cu(
111))
has been explained in detail.
According to our knowledge,
no $\sim 0$ K transient heavy atomic surface alloying and intermixing has been reported 
before.
 We find that
 the Al surface behaves like an ultra-low temperature "atomic trap".
This kind of a mechanism could be general, although we find until now only few
systems (impurity-substrate couple) which show transient inter-layer atomic mixing (TILAM) such as Pt/Cu besides Pt/Al
(also Ir/Al and Au/Al).
Also, no such superdiffusive behavior has been reported yet for intermixing.

 We find that although the mechanism of transient inter-layer atomic mobility is classical, the impinging
impurity particle moves through the potential barrier kinetically, however,
the transition rate does not depend on the external temperature, hence
the process is athermal.
The strong deviation from the Arrhenius law
and the multiply barriered potential energy profile also implies
an unconventional mechanism.

 The understanding of the mechanism could help to explain and classify various athermal atomic transport processes
known in the literature such as superdiffusion on solid surfaces \cite{Michely,Levy,Luedtke,Sule_SUCI}, quantum diffusion \cite{Philibert,Michely,Kramer,Bulou2}, transient enhanced diffusion and intermixing \cite{Sule_JAP07,Abrasonis},
transient cluster burrowing \cite{Sule_cluster}, ultrafast adatom island nucleation
\cite{Sule_SUCI},
coherent ballistic displacement of atoms \cite{Nordlund_Nature},
the cooperative enhancement of surface roughening \cite{Sule_NIMB04b},
the facilitated intermixing near step edges \cite{Lee} or
the superdiffusive mixing in plastically deformed solids \cite{Bellon}.

 During few of these processes no direct external forced conditions haven been applied, or
the athermal process occurs beyond the spatial range of the 
external stimulus \cite{Nordlund_Nature,Sule_JAP07,Sule_SUCI,Abrasonis,Sule_NIMB04b}. 
One can classify them as {\em self-organized facilitated transient} atomic
transport processes which lead to superdiffusion 
due to still unknown or not clearly established reasons.
Until now no considerable effort has been made to understand the driving force of such athermal
transient atomic transport processes.
This piece of a work could help in resolving the mechanism of unconventional atomic transport
which could be widespread in nature.

  The deposited particle gains kinetic energy during local acceleration (its
initial kinetic energy is zero)
towards the surface of the substrate and the arrival energy of few eV is eligible
to overcome the barrier height of the inter-layer transport of interdiffusion.
The impinging Pt atom induces the local heating up (melting) of the Al substrate (heat spike)
and the occurrence of few transient Al atoms.
The impinging impurity atom is driven through a multiply barriered reaction pathway.
The process is purely kinetically governed: transient uppermost layer Al atoms
give downward momentum to the approaching Pt atoms which in turn is injected to the
Al bulk. 

 The mechanism of TILAM is similar to that of the facilitated passive transport of molecules
through a lipid cell membrane. In both cases the process does not
require extra energy in contrast to other passive transport processes in which thermodynamic bias
or concentration gradient drives diffusion.
During facilitated transport "carriers" at the interface "catalyses" diffusion.
In our case few hyperthermal Al atoms promote the process which move coherently
out-of-plane and kicking down the Pt atom when returning back to their equilibrium
position.
During the course of TILAM, 4 transient Al atoms facilitate  
the channeling of the Pt atom through the topmost layer of Al(111).
TILAM is possible due to the delicate interplay between the kinetic and vibrational
degrees of freedom
of the impurity/host couple which allows transient inter-layer atomic transport at $\sim 0$ K.

  These results together with a previous report \cite{Sule_cluster}
indicate that enhanced heavy particle (including heavy atoms and atomic clusters)
bulk mobility could occur in few impurity/substrate couples due to an unprecentended mechanism.
Surprisingly, transient negative atomic mobility is also possible with TILAM: the impinging
particle can move through the top layer of Al(111) against a repulsive surface potential.
The exotic behavior of various impurity/host couples is shown to be largely dependent on
the atomic mass anisotropy of the system while insensitive to the
strength of the heteronuclear interaction.

{\scriptsize
This work is supported by the OTKA grants F037710
and K-68312
from the Hungarian Academy of Sciences.
We wish to thank to K. Nordlund, T. Michely, P. Barna, and to M. Menyh\'ard 
for helpful discussions.
The help of the NKFP project of
3A/071/2004 is also acknowledged.
The work has been performed partly under the project
HPC-EUROPA (RII3-CT-2003-506079).
}

\end{document}